# Study of a new simulation algorithm for dynamical quarks on the APE-100 parallel computer

B. Jegerlehner[a] *

[a]Max-Planck-Institut für Physik
Föhringer Ring 6, D-80805 München

First results on the autocorrelation behaviour of a recently proposed fermion algorithm by M. Lüscher are presented and discussed. The occurence of unexpected large autocorrelation times is explained. Possible improvements are discussed.

## 1. INTRODUCTION

The basic ideas of the method are described in [1] and in this volume in [2] and will not be discussed here. This paper uses the notation of [2]. The tests were done in a $4D$ $SU(2)$ local gauge theory with 2 flavours of Wilson quarks with periodic boundary conditions.

## 2. CURRENT IMPLEMENTATION

Recall that the action of the approximating theory is given by

$$S_b[U,\phi] = S_g[U] + \sum_k [|(Q-\mu_k)\phi_k|^2 + \nu_k^2|\phi_k|^2] \quad (1)$$

with

$$Q = \gamma_5(D+m)/[c_M(8+m)], \quad (2)$$

and $D$ the Wilson-Dirac operator. $c_M \geq 1$ is an arbitrary constant to avoid large bosonic autocorrelation times.

The local structure of the interactions allows us to use conventional MC procedures for gauge and bosonic fields. Namely writing (neglecting constant terms)

$$S_b[\phi_{k,x}] = \phi_{k,x}^\dagger A_k \phi_{k,x} + B_{k,x}^\dagger \phi_{k,x} + \text{h.c.} \quad (3)$$

with $A_k$ constant and diagonal (in the chiral representation), one can use the standard gaussian over-relaxation and heatbath methods. For the gauge fields one may write

$$S[U_{\mu,x}] = \text{tr}\left\{U_{\mu,x}\left(-\frac{\beta}{2}F_g + F_{\mu,x}\right)^\dagger\right\}, \quad (4)$$

where $F_g$ are the staples of the gauge action and $F_{\mu,x}$ is the induced action of the bosonic fields (which is again a quaternion, i.e. can be written as a coefficient times an $SU(2)$ matrix). Written this way, one can again apply heatbath [3,4] and the standard overrelaxation methods for the $SU(2)$ (and generally for $SU(N)$) case.

In our implementation, an iteration is made up of one bosonic heatbath sweep, $N_{or}^b$ bosonic over-relaxation sweeps, one gauge heatbath sweep and finally $N_{or}^g$ gauge overrelaxation sweeps. Recall the gauge case, where $N_{or}^g$ can be used to decrease $\tau_{\text{int}}$ and the dynamical critical exponents [5] substantially.

The implementation was done on a Quadrics Q16 machine with 128 nodes. On one node, the algorithm needs for the bosonic over-relaxation step $170\mu s$ per update, field and site and $250\mu s$ for the heatbath and gauge over-relaxation steps. This allowed us to accumulate up to 200K iterations for each data point. The maximum lattice size which can be reasonably simulated (using preconditioning) on such a machine is roughly a $16^4$ lattice with 100 bosonic fields. On this lattice a sweep would need approximately $5s$ with a memory consumption of roughly 115 MW.

---

*Talk presented at the International Symposium on Lattice Field Theory, Sept. 27-Oct. 1, 1994, Bielefeld

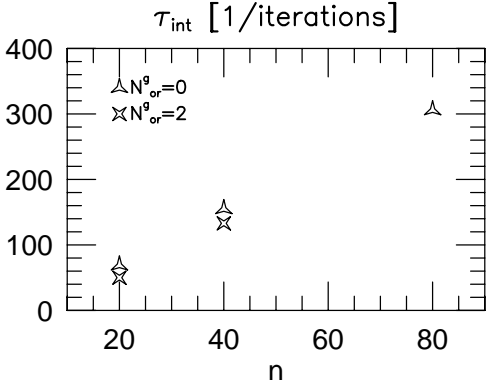

Figure 1. $\tau_{\text{int}}$ against the number of fields for a $4^3 \times 8$ lattice. The bare parameters are $\beta = 1.75$ and $\kappa = 0.165$. $N_{\text{or}}^b$ is 1 in all cases.

## 3. AUTOCORRELATION TIMES

We performed measurements on $4^3 \times 8$ and $6^3 \times 12$ lattices, holding roughly $m_\pi/m_\rho \approx 0.93$ and $m_\pi L \approx 7$. Figure 1 shows the integrated autocorrelation time against the number of the bosonic fields $n$, holding all other parameters fixed. One sees clearly two points: first, $\tau_{\text{int}}$ is roughly proportional to $n$ and secondly, over-relaxing the gauge field does not decrease $\tau_{\text{int}}$ (it even increases $\tau_{\text{int}}$ in units of CPU time linearly). Furthermore, the proportionality factor is rather high.

### 3.1. $n$-behaviour of $\tau_{\text{int}}$

Recall that

$$< A, B > = \frac{1}{2}\text{tr}\left\{AB^\dagger\right\} \tag{5}$$

defines a scalar product on the space of quaternions, so that we may talk about angles and vectors. Let us look at the expectation value of $\text{tr}\{U_{\mu,x} F_{\mu,x}^\dagger\}$. We write here $Q$ as $c_0\gamma_5\{1 - \kappa H\}$ and replace $U_{\mu,x} \to \lambda U_{\mu,x}$ with $\lambda \in \mathbf{R}$ and write

$$\sum_{x,\mu} \text{tr}\left\{U_{\mu,x} F_{\mu,x}^\dagger\right\} = \frac{\partial}{\partial \lambda} S_b \big|_{\lambda=1} - 32(c_0\kappa)^2 \sum_k \phi_k^\dagger \phi_k. \tag{6}$$

The last term comes from the $\lambda^2 U_{\mu,x} U_{\mu,x}^\dagger$ expressions in $Q^2$ and, as we will see, causes the trouble.

Using $\partial Q/\partial \lambda|_{\lambda=1} = Q - c_0\gamma_5$, we write

$$\left\langle \frac{\partial}{\partial \lambda} S_b \big|_{\lambda=1} \right\rangle = \left\langle \text{tr}\left\{ \frac{P'(Q^2)}{P(Q^2)} (Q(Q - c_0\gamma_5) + (Q - c_0\gamma_5)Q) \right\} \right\rangle. \tag{7}$$

$P'(x)/P(x)$ approximates $-1/x$ for $(n, \epsilon) \to (\infty, 0)$ for all polynomials we use, so we can write

$$\left\langle \text{tr}\left\{U_{\mu,x} F_{\mu,x}^\dagger\right\} \right\rangle \approx -4 + \frac{1}{2V}\text{tr}\left\{\frac{c_0\gamma_5}{Q}\right\} - \frac{8(c_0\kappa)^2}{V}\sum_k \left\langle \phi_k^\dagger \phi_k \right\rangle. \tag{8}$$

The first two terms are constant in $n$, but the third one is a sum of $n$ nearly independent variables and thus rises linearly with $n$ (this can be explicitly calculated in the free case). This means that $\|F_{\mu,x}\|$ rises linearly with $n$, which can indeed be observed in the simulation. This explains the unwanted behaviour: after a bosonic update the force induced by the $\phi_k$ fields points roughly in the same direction as the gauge field, with a length proportional to $n$ and an angle of order $1/\sqrt{n}$. The following gauge update allows the gauge field to move only by an angle of the same order. Holding all parameters except $n$ fixed, we may assume, that $\tau_{\text{int}}$ is proportional to the time needed for the gauge field to turn an angle of order 1. Since we perform a random walk this is proportional to $n$, which is exactly what we see. Also multiple updating of the gauge link does not improve the situation since the gauge link will only fluctuate narrowly around the large force, therefore $N_{\text{or}}^g$ has indeed no effect. This situation seems to persist as long as we update the gauge and bosonic degrees of freedom independently.

## 4. IMPROVEMENTS

There are two immediate ideas on how to improve the situation. Firstly, one has to keep $n$ as small as possible and secondly, updating gauge and bosonic fields together may improve the behaviour. The first possibility is already discussed in [2], so we will present an attempt on the second point which is presently being tested.

### 4.1. Modified Updating

One obvious possibility to increase the freedom of the gauge links is to update the link and some bosonic fields together, so that $U_{\mu,x}$ and $F_{\mu,x}$ turn simultaneously. One way to achieve this is to integrate out the bosonic fields at the endpoints of the gauge link and update the link according to the effective action

$$e^{-S_{\text{eff},g}(U_{\mu,x})} = \int \prod_k d\phi_{k,x} d\phi_{k,x+\mu} e^{-S(U_{\mu,x},\phi_{k,x},\phi_{k,x+\mu})} \quad (9)$$

via the heatbath procedure. Since the integrals in $\phi$ are gaussian, one can show that

$$S_{\text{eff},g}(U_{\mu,x}) = \text{tr}\left\{U_{\mu,x}\tilde{F}^\dagger_{\mu,x}\right\}, \quad (10)$$

with $\tilde{F}_{\mu,x}$ of the same form as $F_{\mu,x}$, but slightly more complicated. To fulfill the stability equation $\int dU d\phi p(U,\phi) T(U,\phi \to U',\phi') = p(U',\phi')$, one has to first do a heatbath update $U_{\mu,x} \to_{S_{\text{eff},g}} U'_{\mu,x}$, then update each $\phi_{k,x}$ via heatbath according to

$$e^{-S_{\text{eff},b}(\phi_{k,x})} = \int \prod_k \phi_{k,x+\mu} e^{-S(U'_{\mu,x},\phi_{k,x},\phi_{k,x+\mu})} \quad (11)$$

and finally update $\phi_{k,x+\mu}$ with the standard heatbath procedure. This set of moves is presently being implemented.

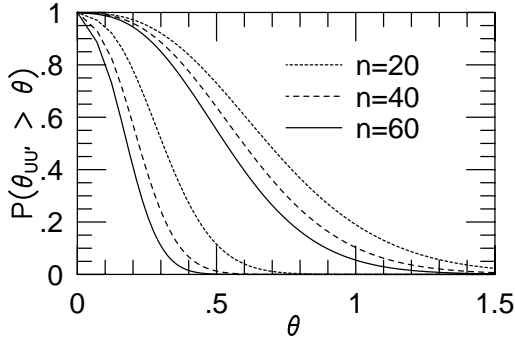

Figure 2. Distribution of the angle between updated and original gauge links using heatbath updating for different $n$. The leftmost curves are done using standard update, the rightmost ones using a combined update.

To estimate the effects of this kind of update we simulated these moves by multiple heatbath updates on the fields $U_{\mu,x}$, $\phi_{k,x}$ and $\phi_{k,x+\mu}$ and measured the probability distribution of the angle between the updated $U'_{\mu,x}$ and $U_{\mu,x}$. The results are shown in Fig. 2. One sees that on one hand the width of the distribution is increased considerably, but that the $n$-dependence is still present. This means that at least the proportionality factor of $n$ in $\tau_{\text{int}}$ will be lowered considerably. This data already indicates that either the $n$-behaviour is changed or the proportionality constant is decreased by at least one to two orders of magnitude smaller than without modification. By implementing this move we expect to improve the efficiency of the algorithm considerably, even taking into account an increased computational cost.

## 5. CONCLUSIONS

While the systematic errors in the approximations seem to be rather uncritical, as stated in [2], the Monte Carlo part of the algorithm needs to be improved further. Some ideas which will help considerably are being implemented, but especially the local structure of the method makes the whole range of existing MC methods (except cluster algorithms) applicable.

## 6. ACKNOLEDGMENTS

I would like to thank M. Lüscher, K. Jansen, R. Sommer and H. Simma for helpful discussions and ideas.